\documentclass[aps,prd,
floatfix,nofootinbib,groupedaddress,showpacs]{revtex4-1}

\usepackage{amsmath}
\usepackage{amssymb}
\usepackage{graphicx}

\usepackage{latexsym}
\usepackage{array}
\usepackage{verbatim}

\usepackage{rotating}
\usepackage{color}

\usepackage[utf8]{inputenc}

\def\lsim{\mathrel{\raise.3ex\hbox{$<$\kern-.75em\lower1ex\hbox{$\sim$}}}}
\def\gsim{\mathrel{\raise.3ex\hbox{$>$\kern-.75em\lower1ex\hbox{$\sim$}}}}

\newcommand{\be}{\begin{equation}}
\newcommand{\ee}{\end{equation}}

\newlength{\absize}
\def\lsim{\mathrel{\rlap{\raise 2.5pt \hbox{$<$}}\lower 2.5pt
\hbox{$\sim$}}}

\allowdisplaybreaks

\begin{document}


\title{Probing $Z$-$Z'$ mixing with ATLAS and CMS resonant diboson  production data \\
at the LHC at $\sqrt{s}=13$ TeV }

\author{P. Osland}
\email{Per.Osland@uib.no}
\affiliation{Department of Physics and Technology,
University of Bergen, Postboks 7803, N-5020 Bergen, Norway}
\author{A.~A. Pankov}
\email{pankov@ictp.it}
\affiliation{The Abdus Salam ICTP
Affiliated Centre, Technical University
of Gomel, 246746 Gomel, Belarus}
\affiliation{Institute for Nuclear Problems, Belarusian State University,
220030 Minsk, Belarus}
\author{A.~V. Tsytrinov}
\email{tsytrin@rambler.ru}
\affiliation{The Abdus Salam ICTP
Affiliated Centre, Technical University
of Gomel, 246746 Gomel, Belarus}
\date{\today}

\begin{abstract}
 The study of electroweak boson pair production provides a powerful test
of the spontaneously broken gauge symmetry of the Standard Model
(SM) and can be used to search for new phenomena beyond the SM.
Extra neutral vector bosons $Z'$ decaying to charged gauge vector
boson pairs $W^+W^-$ are predicted in many scenarios of new
physics, including models with an extended gauge sector. The
diboson production  allows to place stringent constraints on the
$Z$-$Z'$ mixing factor $\xi$ and $Z'$ mass, $M_{Z'}$. We present
the $Z'$ exclusion region in the $\xi-M_{Z'}$ plane for the first
time by using data comprised of $pp$ collisions at $\sqrt{s}=13$
TeV and recorded by the ATLAS and CMS detectors at the CERN LHC,
with integrated luminosities of 36.1 and 35.9 fb$^{-1}$,
respectively. The exclusion region has been significantly extended
compared to that obtained from the previous analysis performed
with Tevatron data as well as with LHC data collected at 7 and 8
TeV. Also, we found that these constraints on the $Z$-$Z'$ mixing
factor are more  severe than those derived from the global
analysis of electroweak data. Further improvement on the
constraining of this mixing can be achieved from the analysis of
data to be collected at higher luminosity.
\end{abstract}

\maketitle

\section{Introduction}\label{sec:I}

Many  new physics (NP) scenarios beyond the SM
\cite{Olive:2016xmw}, including superstring and
left-right-symmetric models, predict the existence of new neutral
and charged gauge bosons, which might be light enough to be
accessible at current and/or future
colliders~\cite{Langacker:2008yv}. The search for these new
neutral $Z^{\prime}$ and charged $W^\prime$ gauge bosons is an
important aspect of the experimental physics program of
high-energy colliders. In this note we concentrate  on
the former one.

Present limits from direct production at the LHC and  virtual
effects at LEP, through interference or mixing with the $Z$ boson,
imply that any new $Z^{\prime}$ boson is rather heavy and mixes
very little with the $Z$ boson. Depending on the considered
theoretical model, $Z^{\prime}$ masses of the order of
4.5~TeV \cite{Aaboud:2017buh, CMS:2015nhc} and $Z$-$Z^{\prime}$
mixing angles at the level of a few per mil are
excluded~\cite{Erler:2009jh} (see also
\cite{Aaltonen:2010ws,Andreev:2012zza}). The mixing angle is
strongly constrained by very high-precision experiments at LEP and
the SLC \cite{ALEPH:2005ab}. They include measurements from the
$Z$  line shape, from the leptonic branching ratios normalized to
the total hadronic $Z$ decay width as well as from leptonic
forward-backward asymmetries. A $Z^{\prime}$ boson, if lighter
than about 5 TeV, could be discovered at the LHC
\cite{Godfrey:2013eta,Dittmar:2003ir} with $\sqrt{s}=14$ TeV in
the Drell-Yan (DY) process $pp \to Z' \to \ell^+ \ell^-+X$
with $\ell=e, \mu$.

After the discovery of a $Z^{\prime}$ boson at the LHC via the DY
process, some diagnostics of its couplings and $Z$-$Z^{\prime}$
mixing needs to be performed in order to identify the underlying
theoretical framework. In this note we investigate the
implications of the ATLAS \cite{ATLAS:2017xvp} and CMS
\cite{CMS:2017skt} data in the diboson channel
\begin{equation}\label{procWW}
pp \to W^+ W^-+X
\end{equation}
to probe the $Z'$ boson that arises, e.g.\ in a popular model
with extended gauge sector proposed in \cite{Altarelli:1989ff}.
The analysis is based on $pp$ collision data at a center-of-mass
energy $\sqrt{s}=13$ TeV, collected by the ATLAS (36.1 fb$^{-1}$)
and CMS ($35.9~\text{fb}^{-1}$) experiments at the LHC. In
particular, the data is used to probe the $Z$-$Z^{\prime}$ mixing.
 In ATLAS $W^+W^-$ events are reconstructed via their semileptonic decays of
the $W$'s where one $W$ boson decays into a charged lepton
($l=e,\mu$) and a neutrino, and the other into two jets
\cite{ATLAS:2017xvp}, while in CMS $W$ bosons decay hadronically
with two reconstructed jets (dijet channel) \cite{CMS:2017skt}.

The $W^\pm$ boson pair production process (\ref{procWW}) is
important for studying the electroweak gauge symmetry. General
properties of the weak gauge bosons are closely related to
electroweak symmetry breaking and the structure of the gauge
sector, like the existence and structure of trilinear couplings.
In addition, the diboson decay modes of the $Z'$ probe the gauge
coupling strength between the new and the standard-model gauge
bosons
\cite{Pankov:1992cy,Andreev:2012cj,Andreev:2014fwa,Andreev:2015nfa}.
Furthermore, the coupling strength strongly influences the decay
branching ratios and the natural widths of such a new gauge boson.
Thus, detailed examination of the process (\ref{procWW}) will both
test the gauge sector of the SM with high accuracy and shed light
on NP that may appear beyond the SM. Here, we examine the
feasibility of observing a $Z^{\prime}$ boson in the $W^\pm$ pair
production process at the LHC, which in contrast to the DY process
is not the principal discovery channel, but can help to understand
the origin of new gauge bosons.

Direct searches for a heavy $WW$~resonance have been performed
at the Tevatron by both the CDF and D0 collaborations. The D0
collaboration explored diboson resonant production up to ${\cal O}(700~\text{GeV})$ using the
$\ell\nu \ell' \nu'$ and $\ell \nu j j$ final states
\cite{Abazov:2010dj}. The CDF collaboration also searched for
resonant $WW$ production in the $e \nu j j$ final state, resulting
in a lower limit on the mass of  $Z'$ and $W'$
bosons~\cite{Aaltonen:2010ws}, excluding masses up to ${\cal O}(900~\text{GeV})$,
depending on the mixing.

The direct $WW$ resonance search by the ATLAS and CMS
Collaborations using, respectively, semileptonic $l\nu jj$
and hadronic final-state events in $pp$
 collision data at 13~TeV set mass limits of ${\cal
   O}(3~\text{TeV})$\footnote{The quoted limit assumes $g_{WWZ'}/g_{WWZ}=(M_W/M_{Z'})^2$.} on such resonances
\cite{ATLAS:2017xvp,CMS:2017skt} (see also \cite{Sirunyan:2016cao,
CMS:2016mwi,TheATLAScollaboration:2015msj}).

In this work, we derive bounds on a possible new neutral spin-1
resonance ($Z^\prime$) from the available ATLAS and CMS data on
$W^+W^-$ pair production \cite{ATLAS:2017xvp,CMS:2017skt}. We
present results as constraints on the relevant $Z$-$Z^{\prime}$
mixing factor introduced in Sect.~II and on the $M_{Z^\prime}$
mass.

The paper is organized as follows.
In Section~II, we briefly review a model involving an additional
$Z^\prime$ boson and emphasize the role of $Z$-$Z^\prime$ mixing
in the process (\ref{procWW}). We give expressions for basic
observables (cross sections) of the process under consideration at
parton and hadron levels.  We also discuss achievable constraints
on $Z^\prime$ model parameters from different experiments
underlining the role of the LHC in substantially improving the current
limits on the $Z$-$Z^\prime$ mixing. Section~III presents some
concluding remarks.

\section{Cross section and constraints on $Z$-$Z'$ mixing}

There are many theoretical models which predict a $Z'$ with mass
possibly in the TeV range. We will consider a NP  model
where $Z^\prime$'s interact with light quarks and charged gauge
bosons via their mixing with the SM $Z$ assuming that the
$Z^\prime$ couplings exhibit the same Lorentz structure as those
of the SM. In particular, in the present analysis we will focus
on a gauge boson of the ``sequential standard model'' (SSM). In
the simple reference model described in
\cite{Altarelli:1989ff,Benchekroun:2001je}, the couplings of the
$Z'$ boson to fermions (quarks, leptons) and $W$ bosons are a
direct transcription of the corresponding standard-model
couplings. Note that such a $Z'$ boson is not expected in the
context of gauge theories unless it has additional couplings to
exotic fermions. However, it serves as a useful reference case
when comparing constraints from various sources. It could also
play the role of an excited state of the ordinary $Z$ in models of
compositeness or with extra dimensions at the weak scale.

In many extended gauge models, while the couplings
to fermions are not much different from those of the SM, the $Z'WW$
coupling is substantially suppressed with respect to that of the SM. In
fact, in an extended gauge model  the standard-model trilinear
gauge boson coupling strength, $g_{WWZ}$ ($=\cot\theta_W$), is
replaced by $g_{WWZ} \rightarrow \xi\cdot g_{WWZ}$, where $\xi =
{\cal C}\cdot(M_{W}/M_{Z'})^{2}$ is the mixing factor and ${\cal
C}$ the coupling strength scaling factor \cite{Langacker:1991pg}. We will set cross
section limits on such $Z'_{\rm SSM}$ as a function of the mass
$M_{Z'}$ and $\xi$. One should note that most $Z'$ search results
report mass limits along the $\xi = (M_{W}/M_{Z'})^{2}$ line
(${\cal C}=1$ is referred to as ``reference model'') and we have also
done so for comparison.

The differential cross section for $Z^{\prime}$ production
in the process (\ref{procWW}) from initial quark-antiquark states
can be written as
\begin{eqnarray}
 \frac{d\sigma^{Z^\prime}}{dM\,dy\,dz}
 = K \frac{2 M}{s}
\sum_q [f_{q|P_1}(\xi_1)f_{\bar q|P_2}(\xi_2) + f_{\bar
q|P_1}(\xi_1)f_{q|P_2}(\xi_2)]\, \frac{d\hat \sigma_{q \bar
q}^{Z^\prime}}{dz}. \label{dsigma}
\end{eqnarray}
Here, $s$ denotes the proton-proton center-of-mass energy squared,
$z\equiv\cos\theta$, with $\theta$ the $W^-$-boson--quark angle in
the $W^+W^-$ center-of-mass frame and $y$ is the diboson rapidity.
Furthermore, $f_{q|P_1}(\xi_{1},M)$ and $f_{\bar
q|P_2}(\xi_{2},M)$ are parton distribution functions for the
protons $P_1$ and $P_2$, respectively, with $\xi_{1,2}=(M/\sqrt
s)\exp(\pm y)$ the parton fractional momenta. Finally, ${d\hat
\sigma_{q \bar q}^{Z^\prime}}/{dz}$ are the partonic differential
cross sections. In~(\ref{dsigma}), the $K$ factor accounts for
higher-order QCD contributions.
 For numerical computation, we use CTEQ-6L1 parton distributions
\cite{Pumplin:2002vw}. Our estimates will be at the Born level,
hence the factorisation scale $\mu_{\rm F}$ enters solely through
the parton distribution functions, as the parton-level cross
section at this order does not depend on $\mu_{\rm F}$. As regards
the scale dependence of the parton distributions we choose for the
factorization scale the $W^+W^-$ invariant mass, $\mu_{\rm
F}^2=M^2=\hat{s}$, with $\hat{s}=\xi_1 \,\xi_2\,s$  the parton
subprocess c.m.\ energy squared. The obtained constraints
presented in the following are not significantly modified when
$\mu_{\rm F}$ is varied from $\mu_{\rm F}/2$ to $2\mu_{\rm F}.$

The cross section for the narrow $Z'$ state production and
subsequent decay into a $W^+W^-$ pair needed in order to estimate
the expected number of $Z'$ events, $N^{Z^\prime}$, is derived
from (\ref{dsigma}) by integrating the right-hand-side over $z$,
over the rapidity of the $W^\pm$-pair $y$ and invariant mass $M$
around the resonance peak $(M_R-\Delta M/2,$ $ M_R+\Delta M/2)$:
\begin{equation}
\sigma^{Z^\prime}{(pp\to W^+W^- + X)}  =\int_{M_{R}-\Delta
M/2}^{M_{R}+\Delta M/2}d M \int_{-Y}^{Y}d y
\int_{-z_{\text{cut}}}^{z_\text{cut}}d
z\frac{d\sigma^{Z^\prime}}{d M\, d y\, d z}\;, \label{TotCr}
\end{equation}
where the phase space can be found, e.g. in
\cite{Andreev:2014fwa}. Using Eq.~(\ref{TotCr}), the number of
signal events for a narrow $Z'$ resonance state can be written as
follows
\begin{equation}
N^{Z^\prime}= {\cal L}\cdot\varepsilon\cdot
\sigma^{Z^\prime}{(pp\to W^+W^- + X)} \equiv {\cal
L}\cdot\varepsilon\cdot A_{WW}\cdot \sigma(pp\to Z') \times
\text{Br}(Z' \to W^+W^-).
 \label{signal}
\end{equation}
Here, ${\cal L}$ denotes the integrated luminosity, and
the overall kinematic and geometric acceptance times trigger,
reconstruction and selection efficiencies,
$A_{WW}\times\varepsilon$, is defined as the number of signal events
passing the full event selection divided by the number of
generated events \cite{Aad:2012nev,ATLAS:2012mec}. Finally,
$\sigma(pp\to Z') \times \text{Br}(Z' \to W^+W^-)$ is the
(theoretical) total production cross section times branching ratio
extrapolated to the total phase space.

The differential cross section for the processes $q\bar{q}\to
Z'_{\rm SSM}\to W^+W^-$, averaged over quark colors, can be
written as \cite{Andreev:2014fwa}
\begin{align}
 &\frac{d\hat{\sigma}^{Z'}_{q \bar q}}{d \cos\theta}
 = \frac{1}{3}\,\frac{\pi\alpha^2 \cot^2\theta_W}{16 }
\left(v_{f}^2 + a_{f}^2\right)\, \frac{\hat{s}}
{\left(\hat{s} - M_{Z'}^2\right)^2 + M_{Z'}^2\Gamma_{Z'}^2}  \nonumber \\
& \times  \xi^2\beta_W^3 \left(\frac{\hat{s}^2}{M_W^4}
\sin^2\theta +
4\frac{\hat{s}}{M_W^2}(4-\sin^2\theta)+12\sin^2\theta\right),
\label{xsection2}
\end{align}
where $v_f=(T_{3,f}-2Q_f\hskip 2pt s_W^2)/(2s_Wc_W)$,
$a_f=T_{3,f}/(2s_Wc_W)$. Finally, $M_{Z'}$ and $\Gamma_{Z'}$ denote the mass
and total width of the $Z'$ boson.

In the calculation of the total width $\Gamma_{Z'}$ we included
the following channels: $Z'\to f\bar f$, $W^+W^-$, and $ZH$
\cite{Barger:2009xg}, where $H$ is the SM Higgs boson and $f$ are
the SM fermions ($f=l,\nu,q$). The total width $\Gamma_{Z'}$ of
the $Z'$ boson can be written as  follows:
\begin{equation}\label{gamma2}
\Gamma_{Z'} = \sum_f \Gamma_{Z'}^{ff} + \Gamma_{Z'}^{WW} +
\Gamma_{Z'}^{ZH}.
\end{equation}
The presence of the two last decay channels,  which are often neglected, 
is due to $Z$-$Z'$ mixing. However for large
$Z'$ masses there is an enhancement that cancels the suppression
due to the tiny $Z$-$Z'$ mixing parameter $\xi$ \cite{Salvioni:2009mt}.
Notice that for all $M_{Z'}$ values of interest for LHC the width
of the $Z'_{\rm SSM}$ boson is considerably smaller than the mass
resolution $\Delta M$.

The expression for the partial width of the $Z'\to W^+W^-$ decay
channel can be written as \cite{Altarelli:1989ff}:
\begin{equation}
\Gamma_{Z'}^{WW}=\frac{\alpha}{48}\cot^2\theta_W\, M_{Z'}
\left(\frac{M_{Z'}}{M_W}\right)^4\left(1-4\,\frac{M_W^2}{M_{Z'}^2}\right)^{3/2}
\left[ 1+20 \left(\frac{M_W}{M_{Z'}}\right)^2 + 12
\left(\frac{M_W}{M_{Z'}}\right)^4\right]\xi^2. \label{GammaWW}
\end{equation}
The dominant term in the second line of Eq.~(\ref{xsection2}), for
$M^2\gg M_W^2$, is proportional to $(M/M_W)^4\sin^2\theta$ and
corresponds to the production of longitudinally polarized $W$'s,
$Z'\to W^+_LW^-_L$. This strong dependence on the invariant mass
results in a very steep growth of the cross section with energy
and therefore a substantial increase of the cross section
sensitivity to $Z$-$Z'$ mixing at high $M$. In its turn, for a
fixed mixing factor $\xi$ and at large $M_{Z'}$ where
$\Gamma_{Z'}^{WW}$ dominates over $\sum_f \Gamma_{Z'}^{ff}$ and
$\Gamma_{Z'}^{ZH}$ the total width increases very rapidly with the
mass $M_{Z'}$ because of the quintic dependence on the $Z'$ mass of
the $W^+W^-$ mode as shown in Eq.~(\ref{GammaWW})
\cite{Altarelli:1989ff}. In this case, the $W^+W^-$ mode becomes
dominant and $\text{Br}(Z' \to W^+W^-)\to 1$, while the fermionic
decay channels are increasingly suppressed. While the Equivalence Theorem 
\cite{Cornwall:1974km,Lee:1977yc}
might suggest a value for $\text{Br}(Z' \to ZH)$ comparable to $\text{Br}(Z' \to W^+W^-)$,
we note that this is inhibited by the vanishing SU(2) structure constant for
the $ZZZ$ coupling.

\begin{figure}[!htb]
\refstepcounter{figure}
\label{fig1}
\addtocounter{figure}{-1}
\begin{center}
\vspace*{-4mm}
\includegraphics[scale=0.51]{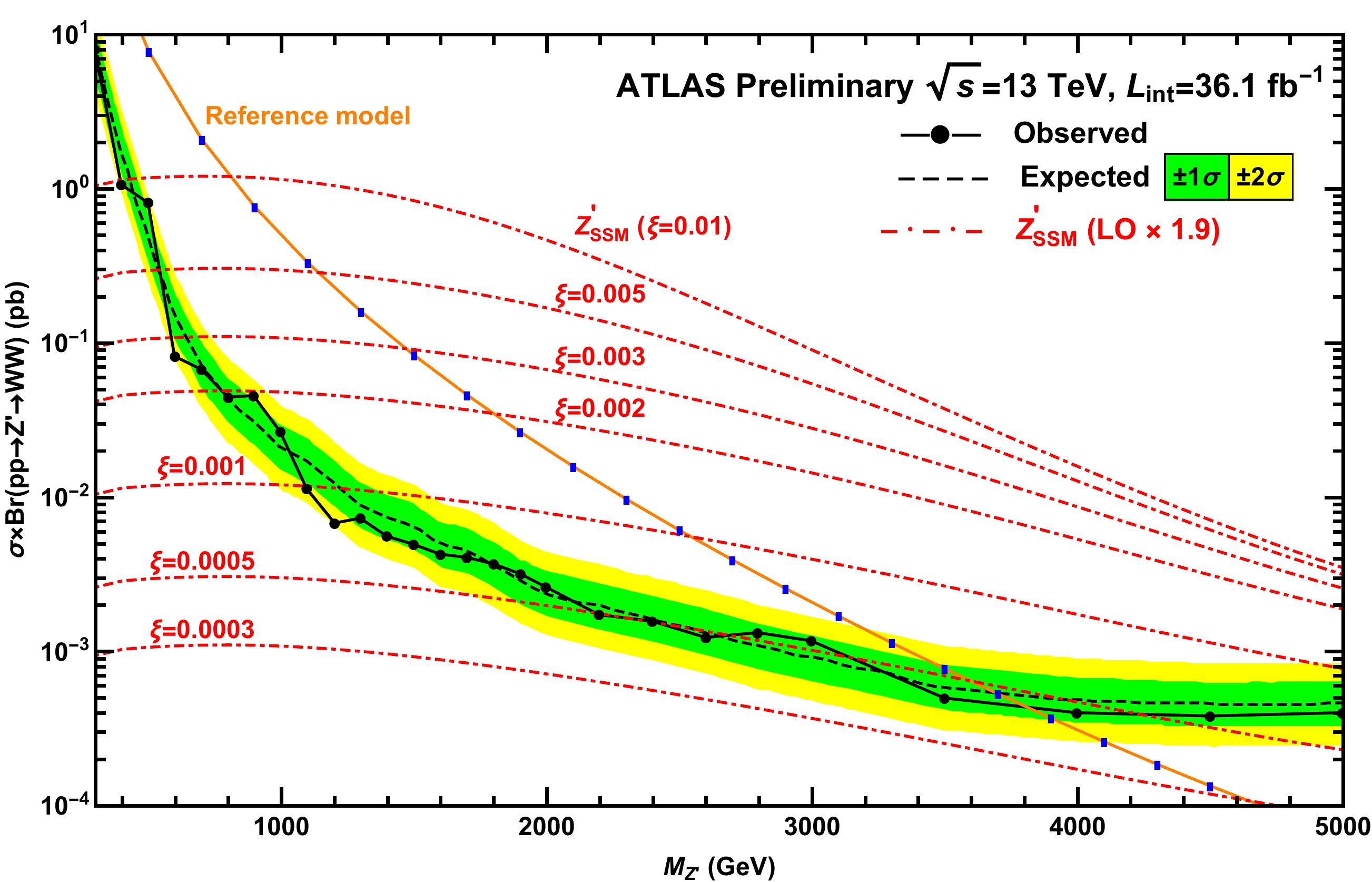}
\includegraphics[scale=0.51]{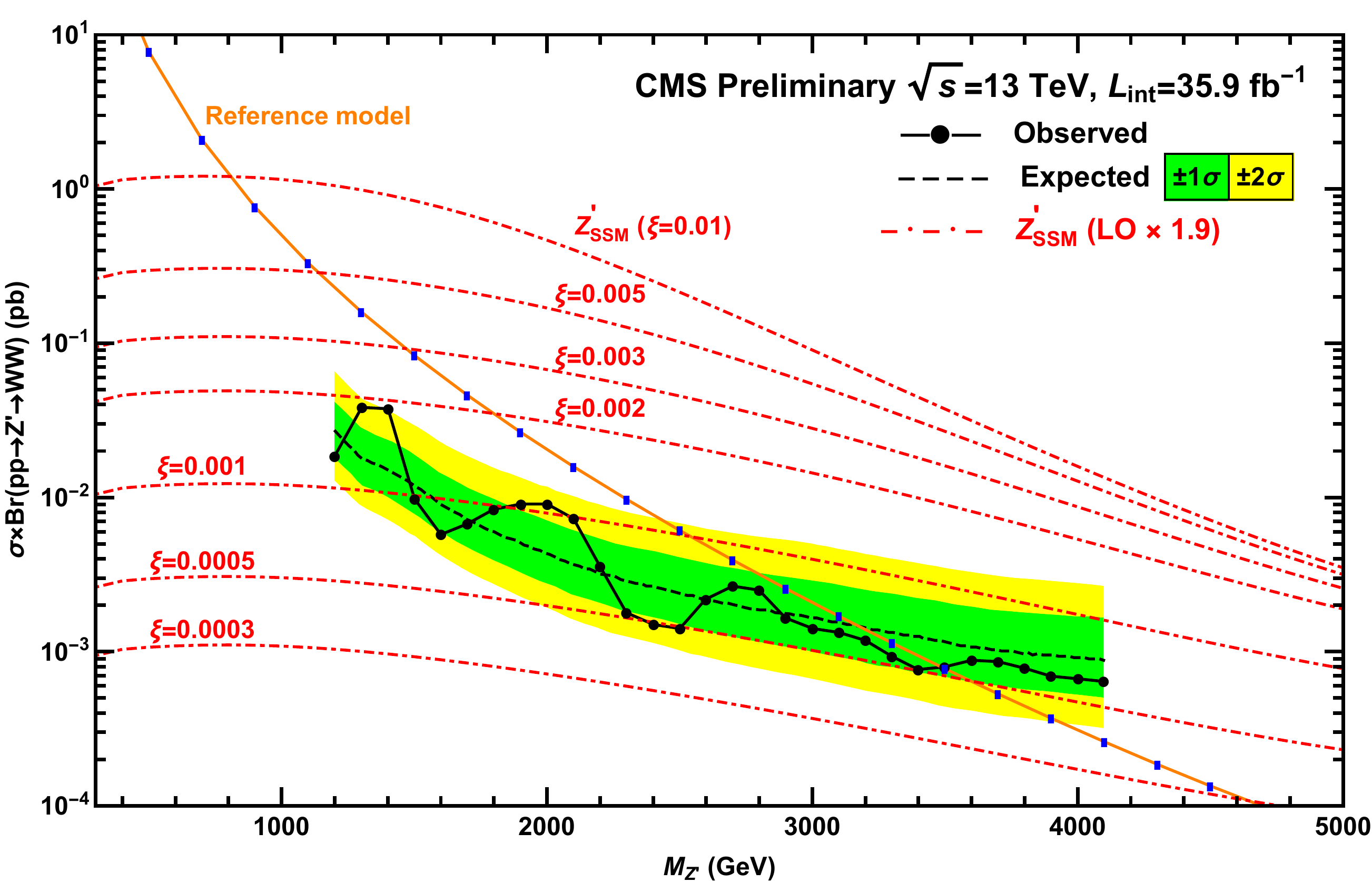}
\end{center}
\caption{Observed and expected $95\%$ C.L. upper limits on the
production cross section times the branching fraction for $Z'\to
W^+W^-$ as a function of $Z'$ mass, $M_{Z'}$. Theoretical
production cross sections $\sigma\times Br(Z'\to W^+W^-)$ for
$Z'_{\rm SSM}$ and reference model are calculated from PYHTHIA
6.409 \cite{Sjostrand:2006za} with a $K$-factor of 1.9, and given
by dash-dotted curves. Labels attached to the  curves for the
$Z'_{\rm SSM}$ cross section correspond to the considered mixing
factor $\xi$. Upper panel: ATLAS data for $36.1~\text{fb}^{-1}$
\cite{ATLAS:2017xvp}, lower panel: CMS data for
$35.9~\text{fb}^{-1}$ \cite{CMS:2017skt}. }
\end{figure}

Further contributions of decays involving Higgs and/or gauge
bosons and supersymmetric partners (including sfermions), which
are not accounted for in (\ref{gamma2}), could increase
$\Gamma_{Z'}$ by a model-dependent amount,  as large as 50\%
\cite{Pankov:1992cy}. In this case, $\Gamma_{Z^\prime}$ would be
larger, with a corresponding suppression in the branching ratio to
$W^\pm$, and the bounds from LHC (and their ability for observing
the $Z$-$Z'$ mixing effect) would be weaker.
\begin{figure}[htb]
\begin{center}
\includegraphics[scale=0.44]{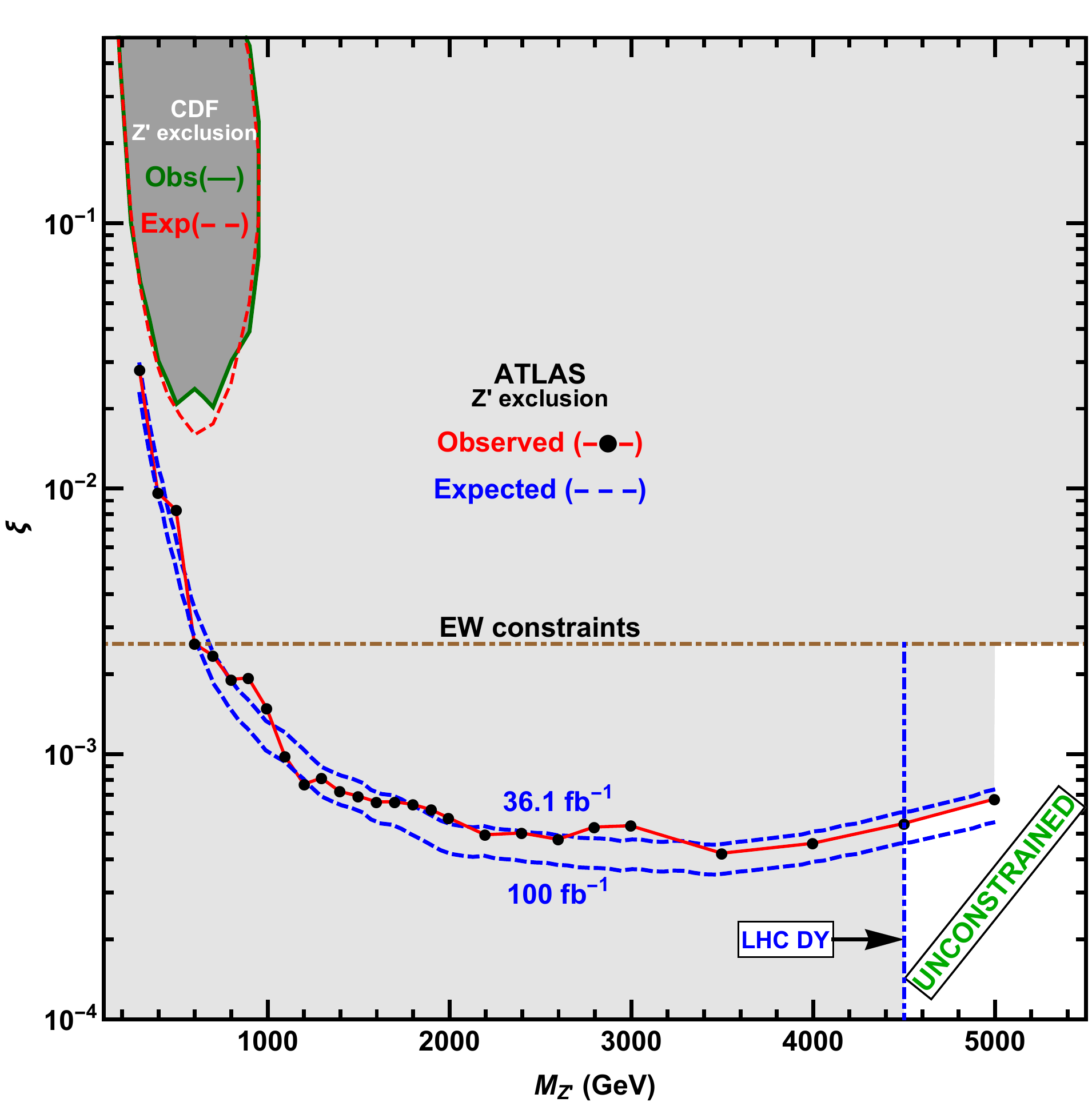}
\includegraphics[scale=0.44]{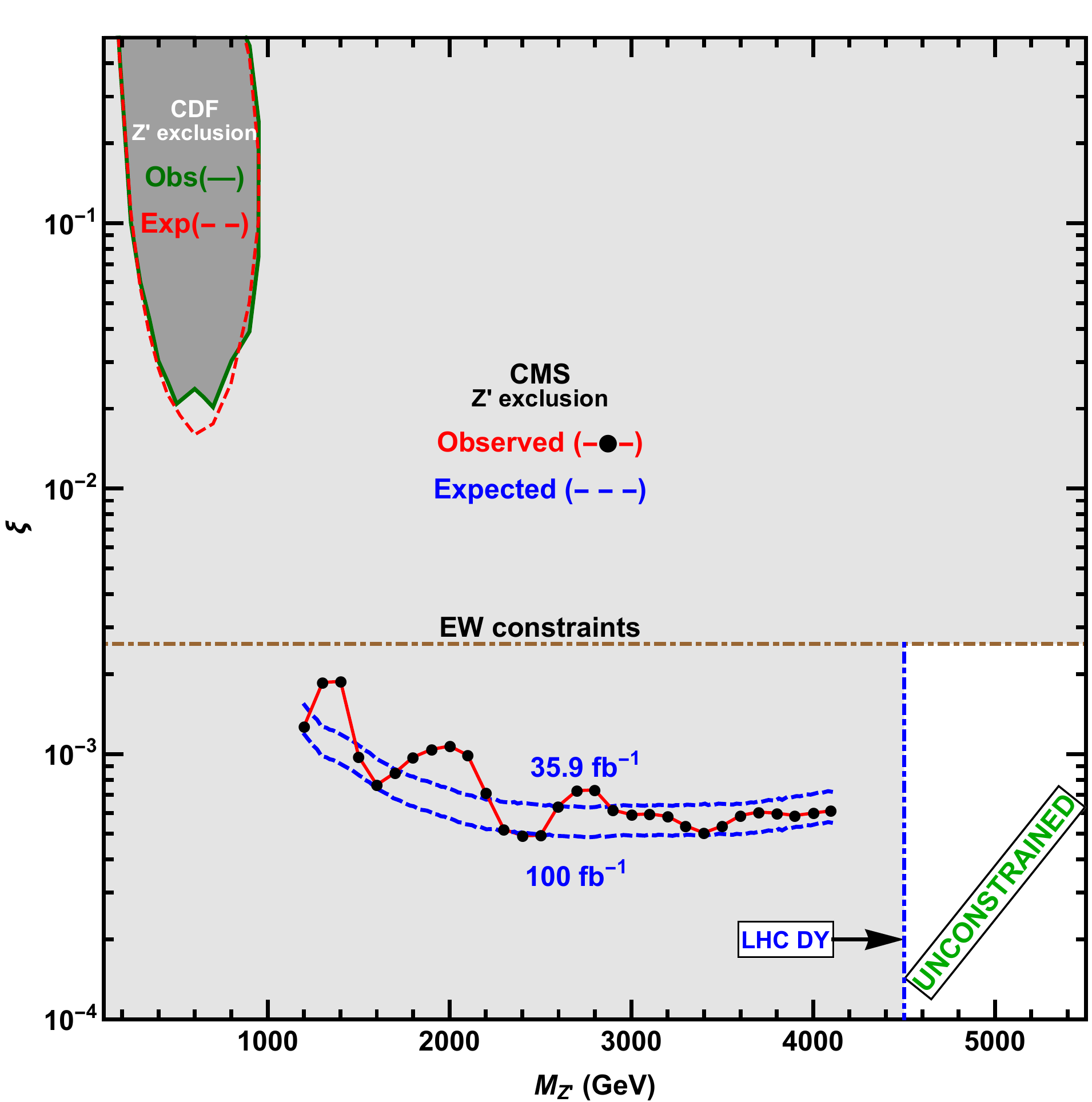}
\end{center}
\caption{$Z'$ exclusion regions in the two-dimensional plane of
($M_{Z'}$, $\xi$) obtained from CDF (Tevatron)
\cite{Aaltonen:2010ws}, precision electroweak (EW) data
\cite{Erler:2009jh} and preliminary LHC data as analyzed here. The
vertical dot-dashed line corresponds to the $Z'_{\rm SSM}$ mass
constraints obtained from the DY process at the LHC 
\cite{Aaboud:2017buh,CMS:2015nhc}. Left panel:
ATLAS data for $36.1~\text{fb}^{-1}$ \cite{ATLAS:2017xvp}, right
panel: CMS data for $35.9~\text{fb}^{-1}$ \cite{CMS:2017skt}.
Exclusion plots with $100~\text{fb}^{-1}$ of data correspond to an
extrapolation of the expected sensitivity.} \label{fig2}
\end{figure}

In our early study \cite{Andreev:2012cj}, to gain some
approximate understanding of the acceptances for signal and
background at different values of the invariant mass $M$ of the
$W^+W^-$ pair and $Z'$ mass, we performed a simple study as
follows. To estimate the 2$\sigma$ constraints on the $Z$-$Z'$ mixing
parameter at the LHC, we compared the events due to a $Z'$ signal to
the events from the SM background in a 3$\%$ interval around the
relevant values of the diboson $W^+W^-$ invariant mass. This should
be compatible with the expected energy resolution and with the
fact that $\Gamma_{Z'} / M_{Z'} \sim 3.0-3.5 \%$. We then required
the signal events to be at least a $2\sigma$ fluctuation over the
expected background, and in any case more than 3. This rough
statistical analysis, as a preliminary stage, was enough to get an
approximate answer to the questions we wanted to address.

Here, we are making a more careful analysis, employing the most
recent measurements of diboson processes provided by the
experimental collaborations ATLAS and CMS, which have control of
the information needed to perform it in a more accurate 
way. In particular, for $Z'_{\rm SSM}$ we compute the LHC
$Z'$ production cross-section multiplied by the branching ratio
into two $W$ bosons, $\sigma(pp\to Z') \times {\rm Br}(Z'\to W^+
W^-)$, as a function of two parameters ($M_{Z'}$, $\xi$), and
compare it with the limits established by the ATLAS and CMS
experiments.

ATLAS \cite{ATLAS:2017xvp} and  CMS  \cite{CMS:2017skt} analyzed
the $W^+W^-$ production in process (\ref{procWW}) through the
semileptonic and hadronic final states, respectively. Our strategy
in the present analysis is to use the SM backgrounds that have
been carefully evaluated by the experimental collaborations and we
simulate only the $Z^\prime$ signal. Fig.~\ref{fig1} shows the
observed and expected $95\%$ C.L. upper limits on the production
cross section times the branching fraction for $Z'\to W^+W^-$ as a
function of $Z'$ mass, $M_{Z'}$. The data analyzed comprises pp
collisions at $\sqrt{s}=13$ TeV, recorded by the ATLAS (36.1
fb$^{-1}$) and CMS (35.9 fb$^{-1}$) detectors at the LHC
\cite{ATLAS:2017xvp,CMS:2017skt}. The inner (green) and outer
(yellow) bands around the expected limits represent $\pm 1\sigma$
and $\pm 2\sigma$ uncertainties, respectively. Also shown are
theoretical production cross sections $\sigma\times
\text{Br}(Z'\to W^+W^-)$ for $Z'_{\rm SSM}$, calculated from
PYHTHIA 6.409 \cite{Sjostrand:2006za} adapted for such kind of
analysis.\footnote{Contributions from $WW$ fusion are not taken into account, 
since $\text{Br}(Z^\prime\to W^+W^-)$ is restricted to the value 
of 2\%  for $M_{Z'}\leq 3.8$~TeV in the parameter region of interest 
(green and yellow bands shown in Fig.~1a),
and actually below 10\% up to $M_{Z'}$ = 5 TeV. Notice that the reference
model predicts $\text{Br}(Z'\to W^+W^-)\approx 2\%$.}
Higher-order QCD corrections for the SM and $Z'$ boson
cases were estimated using a $K$-factor, for which we adopt a
mass-independent value of 1.9
\cite{Frixione:1993yp,Agarwal:2010sn,Gehrmann:2014fva} (however,
note that also a lower value close to unity has been given
\cite{Bai:2012zza}). These theoretical curves for the cross
sections, in descending order, correspond to values of the
$Z$-$Z'$ mixing factor $\xi$ from 0.01 to 0.0005. The intersection
points of the expected (and observed) upper limits on the
production cross section with these theoretical cross sections for
various $\xi$ give the corresponding lower bounds on ($M_{Z'}$,
$\xi$) displayed in Fig.~\ref{fig2}. The  line with the attached
label ``Reference model'' indicates PYTHIA defaults (except for
the above-mentioned $K$-factor) which is commonly used for mass
exclusion regions.  We found that the expected (observed)
exclusion limits are $M_{Z'}<3.7~(3.8)$ TeV (ATLAS) and
$M_{Z'}<3.2~(3.5)$ TeV (CMS).

In Fig.~\ref{fig2}, we collect limits on the $Z'$ parameters, starting with
the Tevatron studies of diboson $W^+W^-$ pair production
\cite{Aaltonen:2010ws}. The limits on $\xi$ and $M_{Z'}$ at the
Tevatron assume that no decay channels into exotic fermions or
superpartners are open to the $Z'$. Otherwise, the limits would be
moderately weaker. Interestingly, Fig.~\ref{fig2} shows that
at heavy $Z'$ masses, the
limits on $\xi$ obtained from the ATLAS and CMS diboson resonance production
data at the LHC at 13 TeV are stronger than
those derived from the global analysis of the precision
electroweak data \cite{Erler:2009jh}.

Also, here we have extrapolated the experimental sensitivity
curves for higher expected luminosity downwards by a factor of
$1/\sqrt D$ where $D$ is the ratio of the expected integrated
luminosity of 100 fb$^{-1}$ that will be collected at Run~2 by
2018 to the current integrated luminosities of 36.1~fb$^{-1}$ and
35.9~fb$^{-1}$. It is clear that further improvement on the
constraining of this mixing can be achieved from the analysis of
such data.

\section{Concluding remarks}
\label{sect:concl}

If a new $Z'$ boson exists in the mass range $\sim$ 4--5~TeV,
its discovery is possible at the LHC in the Drell--Yan
channel. Moreover, the detection of the $Z'\to W^+W^-$ mode is
eminently possible and would give valuable information on the
$Z$-$Z'$ mixing. This paper presents an analysis of $Z$-$Z'$
mixing in the process of $W$ pair production. The analysis is
based on preliminary pp collision data at a centre-of-mass energy
$\sqrt{s}$ = 13 TeV, collected by the ATLAS and CMS experiments at
the LHC. We analyze the popular $Z'_{\rm SSM}$ model and determine
limits on its mass, $M_{Z'}$, as well as on the $Z$-$Z'$ mixing
(angle) factor, $\xi$. We present the $Z'$ exclusion region in the
$\xi-M_{Z'}$ plane for the first time by using these data. The
exclusion limits represent a large improvement over previously
published results obtained at the Tevatron, and also over precision
electroweak data and results obtained from proton-proton collisions at $\sqrt{s} =
7$ and 8~TeV. These are the most stringent exclusion limits to
date on the $\xi-M_{Z'}$ plane. Further improvement on the
constraining of this mixing can be achieved from the analysis of
data which will be collected at higher luminosity in the near
future at Run~2 of the LHC.

\vspace*{-4mm}

\section*{Acknowledgements}

This research has been partially supported by the
Abdus Salam ICTP (TRIL Programme) and the Belarusian Republican
Foundation for Fundamental Research. The work of PO has been
supported by the Research Council of Norway.



\end{document}